\documentclass[11pt]{article}

\usepackage[]{acl}
\usepackage{multirow}
\usepackage{times}
\usepackage{latexsym}
\usepackage{amsmath}
\usepackage{rotating}
\usepackage{array}
\usepackage{longtable}
\usepackage[T1]{fontenc}

\usepackage[utf8]{inputenc}

\usepackage{microtype}

\usepackage{inconsolata}

\usepackage{graphicx}

\title{Acoustic to Articulatory Inversion of Speech; Data Driven Approaches, Challenges, Applications, and  Future Scope}

\author{
 \textbf{Leena G Pillai\textsuperscript{1,2}}
 \textbf{D. Muhammad Noorul Mubarak\textsuperscript{1}}\\
 \textsuperscript{1}University of Kerala\\
 \textsuperscript{2}Digital University Kerala
\\
 \small{
   \textbf{Correspondence:} \href{leenabelieve@gmail.com}{leenabelieve@gmail.com}
 }
}
\begin{document}
\maketitle
\begin{abstract}
This review is focused on the data-driven approaches applied in different applications of Acoustic-to-Articulatory Inversion  (AAI) of speech. This review paper considered the relevant works published in the last ten years (2011-2021). The selection criteria  considered are (a) type of AAI – Speaker Dependent and Speaker Independent AAI, (b) objective of the work – Articulatory  approximation, Articulatory Feature space selection and Automatic Speech Recognition (ASR), explore the correlation between  acoustic and articulatory features, and framework for Computer-assisted language training, (c) Corpus – Simultaneously recorded  speech (wav) and medical imaging models such as ElectroMagnetic Articulography (EMA), Electropalatography (EPG),  Laryngography, Electroglottography (EGG), X-ray Cineradiography, Ultrasound, and real-time Magnetic Resonance Imaging  (rtMRI), (d) Methods or Models – recent works are considered, and therefore all the works are based on machine learning, (e)  Evaluation – as AAI is a non-linear regression problem, the performance evaluation is mostly done by Correlation Coefficient (CC),  Root Mean Square Error (RMSE), and also considered Mean Square Error (MSE), and Mean Format Error (MFE). The practical  application of the AAI model can provide a better and user-friendly interpretable image feedback system of articulatory positions,  especially tongue movement. Such trajectory feedback system can be used to provide phonetic, language, and speech therapy for  pathological subjects.  
\end{abstract}

\section{Introduction}
Human speech production is a complex process that is stacked with four sequential processes - initiation, phonation, oro-nasal process, and articulation \cite{Giegerich1992}. The air stream that expelled from the lungs is the source of the speech. This air passes through the larynx, where the vocal folds state adds the voiced or voiceless features to the air stream. The velum controls the next stage of the air stream, and its movement is responsible to determines where the air pass through the nasal (nasal sounds-/m/, /n/, etc.) or oral cavity. Articulation is the process that shapes the air stream by chaining the place (where) and manner (how) of articulation. Basically, a phonological feature set, such as +bilabial, +plosive, -voicing, and so on, individually recognizes each phoneme and their phone classification conveys the information about the place and manner of articulation. Therefore, the interesting concern here is to explore the relationship between the articulatory and the acoustic space of the speech signal.

\begin{figure}[h]
  \includegraphics[width=0.46\textwidth]{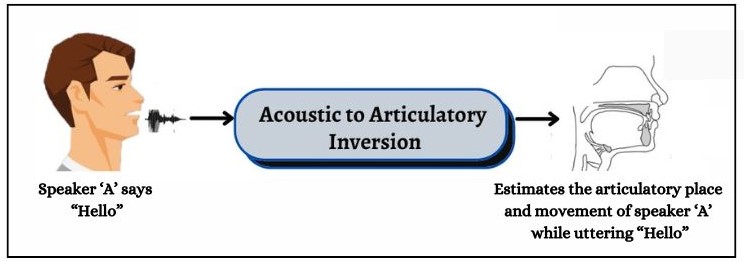}
  \caption{From the Acoustic Features of the speech wave, the AAI estimate how the speech sounds are produced (Place and manner of the articulation).}
  \label{fig:AAI}
  \vspace{-0.2cm}
\end{figure}

The Acoustic-to-Articulatory Inversion (AAI) mapping problem can be defined as a non-linear regression problem that is entitled to inference the articulatory trajectories from the acoustic speech signal \cite{Atal1978, Toda2008, Ananthakrishnan2011}. Acoustic to Articulatory inversion is illustrated in Figure. \ref{fig:AAI}. Many studies were done on this acoustic articulatory correlation and applied in many applications like speech signal analysis and speech synthesis \cite{Ling2009, Steiner2012}, Automatic Speech Recognition \cite{Kirchhoff2002, Krishna2019}, speaker recognition \cite{Hong2020}, speech evaluation of pathological subjects \cite{Moro-Velazquez2021}, diagnosis of depression from speech \cite{Seneviratne2021}, and so on.

The forward problem that is identifying the acoustic features from the speech signal is a well-expanded research area. But the reverse problem, that is, identifying the articulatory features by the acoustic features, is really mystifying researchers over three decades. The main challenge in the AAI problem is its one-to-many nature. An articulatory state always gives a single acoustic comprehension, whereas an acoustic signal is the outcome of multiple articulatory states. As similar articulatory states can utter different acoustic signals, the problem is recognized as non-linear. But the feasible solutions and more enhancements are required in this research area [AAI]. The studies reveal that this accurate articulatory feature extraction and its appropriate representation can be a good aid for pathological language and speech development \cite{Li2019, Jones2017}.

This work focuses on the review of the selected research papers published from 2011 to 2021. The selection criteria are based on publication time, objective, corpus, features considered, machine learning models, and quantitative outcome of the work.

\section{Objective}
This work focused on the study and discussion of the interesting area of research called Acoustic-to-Articulatory Inversion (AAI) of speech. The work selected is in such a way to familiarize the readers in the concept of AAI, its challenges, application areas, and moreover that the approaches experienced in this area, right from the data source availability, feature extraction methods, modeling, and evaluation methods.

\section{Motivation}

The innovative and effective applications of AAI have high-level relevance in the betterment of society. In many cases, it is not applicable to use medical imaging models to evaluate the articulatory position and movement of the speaker during speech, and such cases we can suggest AAI articulatory feedback system to track the articulatory movements during speech production. Pronunciation, speech, and language learning is a challenging task for the hearing-impaired community. A self-explanatory, visual feedback system can make them understand easily and clearly the trajectory positions they are currently following and the actually required position they should practice to achieve.

\section{Literature Review}

Several studies reveal that the visual feedback of the articulatory position of the speaker is a recommended aid for speech therapy and language training \cite{Byun2014, Roxburgh2015, Cleland2015}. Initial works on AAI was concentrated on speaker-dependent approaches (SD-AAI), which is based on codebook \cite{Hogden1996, Ouni1999}. The Suzuki, S., Okadome T, \& Honda, M explored in their work that the performance of the AAI is dramatically enhanced by introducing dynamic constraints \cite{Suzuki1998}. An inversion based on feed-forward neural network model was proposed by Richmond K, King S, \& Taylor P., \cite{Richmond2003}.

To meet the real-world demands, the inversion system needs to be robust and speaker-independent (SI-AAI). Lately, the availability of simultaneously recorded articulatory and acoustic corpus radically increased the scope of SI-AAI. As the training data corpus availability increased, the traditional models uplifted to data-driven methods such as various Deep Artificial Neural Network frameworks \cite{Cai2018, Seneviratne2019}, Mixture Density Networks \cite{Uria2012, Illa2020}, Recurrent Neural Network \cite{Liu2015, Xie2018_isclsp, Biasutto-Lervat2018}, Convolutional Neural Network \cite{Yu2018, Mannem2019}, and so on.

\section{Approaches towards the solution}
This problem area is mostly following data-driven approach. Now a days, machine learning is the dominating approach for any classification and regression problem. Any problem in machine learning should went through a sequence of activities such as (1) corpus collection and preprocessing, (2) feature extraction, (3) training and testing.

\subsection{Speech - Articulatory Databases}

Several approaches are available to simultaneously capture speech signals and its corresponding articulatory data, listed in Table. \ref{tab:arti}.

\begin{table*}[t!]
    \centering
    \begin{tabular}{|p{4cm}|p{11cm}|}
        \hline
        \textbf{Techniques} & \textbf{Description} \\
        \hline
        Electropalatography (EPG) & During speech, the tongue and palate contacts are recorded by using an artificial palate which embedded with 62 silver electrodes \cite{Hardcastle1972}. \\
        \hline
        Laryngography or Electroglottography (EGG) & Laryngographic signals, high frequencies similar to actual audio, are recorded by two electrodes which is placed around the neck \cite{Baken1992}. \\
        \hline
        Pneumotachography & Used to measure the oral and nasal airflow and might provide additional information for AAI \cite{Miller1993}. \\
        \hline
         X-ray Cineradiography & x-ray filming of trachea while speech production. Not recommended because of radiation \cite{Munhall1995}. \\
        \hline
        ElectroMagnetic Articulography (EMA) & Sensor coils attached to lip, tongue, teeth and velum of the speaker and he/she should wear a helmet to record trajectories \cite{Ryalls2000}. \\
        \hline
        Magnetic Resonance Imaging (MRI) & MRI scanner used to capture the upper airways of the speaker \cite{Narayanan2014}. \\
        \hline
        Ultrasound & Commonly used biofeedback imaging system which is generated by high frequency sound waves \cite{Wei2016}. \\
        \hline
    \end{tabular}
    \caption{Techniques to capture articulatory data during speech}
    \label{tab:arti}
\end{table*}

The articulatory dataset which is parallelly recorded with speech is very limited. The time consumption, requirements of the equipment, and uncomfortableness of attaching such equipment with the human body are the main parameters that restrict from having a large dataset. The commonly used dataset which is available for the AAI work is listed in Table. \ref{tab:dataset}:

\begin{table*}[t!]
    \centering
    \begin{tabular}{|p{1.5cm}|p{6cm}|p{7cm}|}
        \hline
        \textbf{Corpus} & \textbf{Description} & \textbf{Accessible link} \\
        \hline
        MOCHA- TIMIT & 460 British English sentences from a female speaker and corresponding EMA data for articulatory features. & \href{https://data.cstr.ed.ac.uk/mocha/}{https://data.cstr.ed.ac.uk/mocha/} \\
        \hline
         USC & Magnetic Resonance Imaging (MRI), and EMA collected from American male speaker (spontaneous speech). & \href{https://sail.usc.edu/span/usc-timit/}{https://sail.usc.edu/span/usc-timit/} \\
        \hline
        MNGUO & Parallel recordings of acoustic and articulatory (EMA) data of 1263 English utterances. & \href{http://www.mngu0.org/}{http://www.mngu0.org/} \\
        \hline
        TORGO & Articulatoryacoustic -data of dysarthric speakers. & \href{http://www.cs.toronto.edu/~complingweb /data/TORGO/torgo.html}{http://www.cs.toronto.edu/$\sim$complingweb /data/TORGO/torgo.html} \\
        \hline
        XRMB & X-ray Microbeam Database of 46 speakers. & \href{https://vale.app.box.com/s/cfn8hj2puveo 651q54rpIml2mk7moj3h/folder/30415804819}{https://vale.app.box.com/s/cfn8hj2puveo 651q54rpIml2mk7moj3h/folder/30415804819} \\
        \hline
    \end{tabular}
    \caption{Publicly available acoustic-articulatory dataset}
    \label{tab:dataset}
\end{table*}
MOCHA is a commonly used corpus. The preprocessing is mostly do silence removal, noise filtering, and sequential alignment of the acoustic speech and articulatory data. Some domain centered work is not able to work with this common public dataset. Therefore, they required to build their own dataset \cite{Hueber2012, Hueber2013, Xie2016, Fabre2017}.

\subsection{Feature Extraction}

Feature extraction is an important decision in machine learning, as it represents the data at hand. The most suitable representation will accelerate the system's accuracy. AAI problem demands two sets of feature sets: one set represents the acoustic space, and another represents the articulatory space. Acoustic space is well established in the area of Automatic Speech Recognition \cite{Rabiner1993Fundamentals}. Mel Frequency Cepstral Coefficients (MFCC) is widely accepted and uses acoustic space feature extraction techniques. The MFCC and its derivatives provide a set of the subject or speaker independent features. The vocal tract shapes the sound and envelops the features in the short-time power spectrum. The MFCC represents this envelope which means it is directly proportional to the trajectory features. Other acoustic features such as scattering coefficients (SCs) and Discrete Wavelet Coefficients (DWCs), and Logarithm of square Hanning Critical Bank filter banks (LHCB) are used in our considered literature. Recently, applied normalization methods, such as Cepstral Mean and Variance Normalization (CMVN) and Vocal Tract Length Normalization (VTLN) on acoustic features to minimize the speaker-specific information (Table 3).

\begin{figure}[h!]
    \centering
    \includegraphics[width=0.23\textwidth]{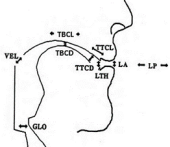} 
    \caption{Tract Variable \cite{Sivaraman2019}}
    \label{fig:TV}
\end{figure}

Articulatory Features (AF) are important in this problem. Each articulation represents within the Tract Variable (TV), which include Lip Aperture (LA), Lip Protrusion (LP), Tongue Dorsum Constriction Location (TDCL), Tongue Dorsum Constriction Degree (TDCD), Tongue Tip Constriction Location (TTCL), Tongue Tip Constriction Degree (TTCD), VELar opening (VEL), Lower Tooth Height (LTH) and GLOttal vibration (GLO) (Figure. \ref{fig:TV}).

For example, if the utterance is made by closing the lips together, then the LA would be close to zero. Most of the work used ElectroMagnetic Articulography (EMA) biofeedback images. The above-mentioned TVs (the whole or some of the features) are derived from EMA, which tracks the trajectories.

The Euclidean Distance between the x, y sensor coordinates of Upper Lip (UL) and Lower Lip (LL) is termed as Lip Aperture (LA) (Eq.\ref{eq:la}):
 \vspace{-0.2cm}
\begin{equation}
    \small
    LA[i] = \sqrt{LL_{x}[i]- (UL_{x}[i]) + LL_{y}[i] - (UL_{y}[i])^{2}}
    \label{eq:la}
\end{equation}

The Lip Protrusion (LP) describes the horizontal Lower Lip (LL) displacement along the x axis of the relavant sensor (Eq.\ref{eq:lp}):

\begin{equation}
    LP[i]=LL_{x}- \underset{m \in utterances}{median}\{LL_{x}[m]\}
    \label{eq:lp}
\end{equation}

The tongue constriction location describes the horizontal x axis displacement of tongue from its median position. Most commonly used tongue constriction locations are Tongue Tip Constriction Location (TTCL), Tongue Dorsum Constriction Location (TDCL), and Tongue Body Constriction Location (TBCL). The calculation method for the above mentioned three tongue constriction location are as follows (Eq.\ref{eq:ttcl}, Eq. \ref{eq:tdcl} and  Eq.\ref{eq:tbcl}):

\begin{equation}
    \small
    TTCL[i]=TT_{x}[i]-\underset{m \in utterances}{median}\{TT_{x}[m]\}
    \label{eq:ttcl}
    \vspace{-0.2cm}
\end{equation}

\begin{equation}
    \small
    TDCL[i]=TD_{x}[i]-\underset{m \in utterances}{median}\{TD_{x}[m]\}
    \label{eq:tdcl}
    \vspace{-0.1cm}
\end{equation}

\begin{equation}
    \small
    TBCL[i]=TB_{x}[i]-\underset{m \in utterances}{median}\{TB_{x}[m]\}
    \label{eq:tbcl}
    \vspace{-0.1cm}
\end{equation}

The minimum distance between the palate and tongue dorsum and tongue tip are defined by Tongue Dorsum Constriction Degree (TDCD) and Tongue Tip Constriction Degree (TTCD).

The 'EigenTongue' (ET) features are extracted from Ultrasound images which enable to encode the trajectory- related maximum information such as hyoid bone, tongue position, and muscles. The features considered in some of the works are given in Table. \ref{tab:features}.

\begin{table*}[h!]
\begin{tabular}{|p{2.5cm}|p{6cm}|p{6.5cm}|}
\hline
\textbf{Ref:} & \multicolumn{2}{c|}{\textbf{Features}} \\
\cline{2-3} & \textbf{Acoustic} & \textbf{Articulatory} \\
\hline
\small{\citealt{Ghosh2011}} & \small{13 dimensional MFCC acoustic features} & \small{5 Tract Variable (LA, LP, jaw opening, tongue tip and tongue body)} \\
\hline
\small{\citealt{Hueber2012}} & \small{12 MFCC features, its delta, and delta-delta features (altogether 39 dimension)}  & \small{AF and Articulatory Phonetic Features (APF) with 30 channels consisting of length, width and depth of 10 critical muscles} \\
\hline
\small{\citealt{Hueber2013}} & \small{The audio signals were parameterized by 25 Mel-Cepstral Coefficients} & \small{6 AF (jaw, tongue tip, blade, dorsum, lower lip, and upper lip) were down sampled to 100 Hz and filtered to 20 Hz by low pass filter} \\
\hline
\small{\citealt{Sudhakar2014}} & \small{39 MFCC with its first and second order derivatives} & \small{14 dimensional Articulatory Features (AF) from EMA (x, y co-ordinates)} \\
\hline
\small{\citealt{Illa2020}} & \small{40dimensional -Line Spectral Frequencies acoustic features with gain value} & \small{12-dimensional AF vector from each 200 Hz sampled EMA data frame} \\
\hline
\small{\citealt{Xie2015}} & \small{39 dimensional MFCC along with delta and delta-delta features} & \small{13dimensional -static AF vector obtained by the coordinates of sensors} \\
\hline
\small{\citealt{Najnin2015}} & \small{39 acoustic feature spaces (MFCC, their energy, velocity, and acceleration)} & \small{14-D AF along with 6-dimensional Tract Variable (TV) from the x, y coordinate of EMA data} \\
\hline
\small{\citealt{Afshan2015}} & \small{39dimensional -MFCC with energy, delta-delta feature vector as acoustic feature} & \small{14-D AF from EMA (x, y coordinates of 7 sensors) \& 6-dimensional Tract Variable (TV) features} \\
\hline
\small{\citealt{Tobing2016}} & \small{24 Mel-cepstral}  & \small{14-dimensional Articulatory Features (EMA) converted to 2- scores} \\
\hline
\small{\citealt{Rudzicz2016}} & \small{39 MFCC (12 CC + energy + delta + delta-delta), 256 wavelet coefficients, and 118 Scattering Coefficients (SC)} & \small{14-dimensional Articulatory Features (EMA data)} \\
\hline
\small{\citealt{Xie2016}} & \small{39dimensional -MFCC acoustic feature} & \small{13-dimensional static EMA feature vector} \\
\hline
\small{\citealt{Wei2016}} & \small{Four spectral formants used as acoustic feature} & \small{30dimension -shared representation of ultrasound image} \\
\hline
\small{\citealt{Sivaraman2016}} & \small{13-dimensional MFCC acoustic features} & \small{6 Tract Variable (TV) articulatory features} \\
\hline
\small{\citealt{Uchida2017}} & \small{24-MFCC from each frame} & \small{14-dimensional AF (7 points X 2 dimensions)} \\
\hline
\small{\citealt{Fabre2017}} & \small{26 MFCC, including static coefficients and their first derivatives} & \small{30-D features from 64 X 64 pixels down sampled and 4096 XI vector transformed ultrasound images. 14-D features from EMA dataset} \\
\hline
\small{\citealt{Porras2019}} & \small{50 acoustic features that includes 25 MFCC features and its first derivatives} & \small{Ultrasound Tongue Imaging (UTI) of 64X64 pixels, 128 EigenTongues (ET) were used as articulatory features} \\
\hline
\small{\citealt{Bozorg2020}} & \small{Mel spectrogram features used for acoustic modeling} & \small{10 articulatory features for kinetic space modeling (6 - tongue movement, 3 lip movement and 1 - jaw movement)} \\
\hline
\small{\citealt{Sun2021}} & \small{MFCC with delta and delta-delta, normalized by CMVN, and VTLN} & \small{12dimensional -Articulatory Features (EMA data) and 4-dimension Vocal Tract (VTs) variable} \\
\hline
\small{\citealt{Maharana2021}} & 3\small{9 dimensional MFCC} & \small{8 dimensional articulatory features (EMA)} \\
\hline
\end{tabular}
\caption{Acoustic \& Articulatory features of the considered works}
\label{tab:features}
\end{table*}

\subsection{Training and Testing}

The technique that can be used for AAI can be classified as Generative models and Discriminative models, which comes in the statistical model. The Generative models are entitled to find the conditional probability $P(Y|X)$ by estimating their prior probability $P(Y)$ and likelihood probability $P(X|Y)$ using the training data. The Posterior Probability $P(Y|X)$ of the Generative model is calculated by Bayes Theorem [e.g., Hidden Markov Model (HMM), Gaussian Mixture Model (GMM)...]. The Discriminative model finds the posterior probabilities by approximating the parameters of $P(Y|X)$ directly from the training data [e.g.. Support Vector Machine (SVM), Deep Neural Network models...)].

As the AAI problem is a nonlinear regression problem in nature, the evaluation is performed by calculating the difference between (error) the estimated or predicted trajectory features from the acoustic features and the actual trajectory features. As discussed before this error calculation is commonly done by CC (Eq.(8)), MSE $(Eq.(9))$, and RMSE (Eq.(10)) (Table 4). The correlation coefficient (CC) defines the strength of the linear relationship between two variables x, y.

\begin{equation}
    CC=\frac{\Sigma(x_{i}-\overline{x})(y_{i}-\overline{y})}{\sqrt{\Sigma(x_{i}-\overline{x})^{2}\Sigma(y_{i}-\overline{y})^{2}}}
    \label{eq:cc}
    \vspace{-0.2cm}
\end{equation}

\begin{equation}
    MSE=\frac{1}{n}\sum_{i=1}^{n}(x_{i}-y_{i})^{2}
    \label{eq:mse}
\end{equation}

\begin{equation}
    RMSE=\sqrt{\frac{{\Sigma_{i=1}^{n}{(x_{i}-y_{i})}^{2}}}{N}}
\end{equation}

Where, $\overline{x}$ is the mean of x, and $\overline{y}$ is the mean of y. N is the number of data points.

The minimum RMSE results in a better model. The implementation of the AF along with acoustic features for classification problems such as Automatic Speech Recognition, Speaker Recognition \cite{Kirchhoff2002, Hong2020}. Table. \ref{tab:out} explicate the methods, and outcome of the work considered.

\begin{table*}[h!]
\centering
\begin{tabular}{|p{2cm}|p{5.5cm}|p{7cm}|}
\hline
\textbf{Ref:} & \textbf{Method/Model} & \textbf{Outcome} \\
\hline
\small{\cite{Ghosh2011}} & \small {The Expectation Maximization (EM) algorithm used for parameter estimation and each frame classified using max-a posteriori (MAP) rule} & \small{The MOCHA female dataset enhanced the accuracy of 81.28\% with acoustic and articulatory (TV) features where the acoustic only was 79.10\%. The USC male dataset also shown 2\% improved accuracy (76.29\%) by including articulatory features} \\
\hline
\small{\cite{Hueber2012}} & \small{ Articulatory Phonetic Inversion (API) find the relation between APF (articulatory) and MFCCs (acoustic) and the data pairs are prepared by Heuristic Learning Algorithm} & \small{The artificial synthesizer generates speech that are closely similar to human speech (70\% - sonorants and 30\% obstruent). APF has shown better overall frame-based phoneme level accuracy (85.6\%) which syllables to improve the recognition accuracy} \\
\hline
\small{\cite{Sudhakar2014}} & \small{GMM with 64 mixtures used for acoustic to articulatory mapping. Sparse smoothing, low pass smoothing, and articulatory features directly from GMM inversion (without smoothing) were compared} & \small{In male and female subject GMM (male-58.86\%, female – 61.70\%) with low pass smoothened feature attained least performance among GMM with original feature (male – 65.47\%, female – 70.68) and GMM with sparsely smoothened feature (male – 65.21\%, female – 70.05\%)} \\
\hline
\small{\cite{Liu2015}} & \small{Another method used is Deep Bidirectional Long Short-Term Memory (DBi-LSTM) in which first 2 layers with 300 units used for feature extraction, the dynamic nature of input learned by 2 Bi-LSTM layers with 100 units} & \small{RMSE is used to measure the performance. In standard input, the DRMDN has shown the better performance (0.948mm) than DBLSTM (0.963mm) and DNN (1.237mm)} \\
\hline
\small{\cite{Xie2015}} & \small{Two streams of independent HMMs, Multiple Regression HMM (MR-HMM), and GVP-HMM were used as the baseline model. Sigmoid DNN and one Deep Mixture Density Network (DMDN) used for inversion} & \small{RMSE used as evaluation criteria. The GVP-HMM model receives the DNN bottleneck features as auxiliary inputs outperformed baseline HMM (0.20mm), MR-HMM (0.16mm), DNN (0.12mm) and DMDN (0.10mm)} \\
\hline
\small{\cite{Najnin2015}} & \small{GRNN is designed with 53 (39 + 14) input layer. 10-fold cross-validation is used to select the smoothing parameter. DBN designed with four hidden layer and each with 300 units} & \small{The RMSE for speech inversion shows that GRNN (TIMIT-1.09mm and MNGU0-0.926mm) performs slightly better than DBN (TIMIT-1.13mm and MNGU0-0.956mm) model} \\
\hline
\small{\cite{Afshan2015}} & \small{HMM adaption is performed by MLLR adaption method. Estimated articulatory features smoothened using low- pass filter. Phone recognition accuracy is evaluated with HMM} & \small{An average 20\% improvement explored in the performance of phone recognition with adapted parameter HMM than non-adaptive} \\
\hline
\small{\cite{Tobing2016}} & \small{Combined the LT with GMM for inversion. The LT-GMM uses the trained model from conventional GMM as initial model. The number of mixture components was varied to 16,32,64 and 128. Parameters are optimized using EM algorithm} & \small{The computational time required to train the LT- GMM is about 15 times slower than conventional GMM. The conventional GMM RMSE declines from 1.9mm to 1.75mm whereas in case of LT-GMM the performance varies from 1.85mm to 1.65mm} \\
\hline
\small{\cite{Rudzicz2016}} & \small{Feature selection done with SVM. AAI experimented with DBNs (with 2 hidden layers) Vs SPN (used to model the observations of HMM)} & \small{The Cerebral Palsy (CP) nasal sounds are reconstructed more accurately using DBN (0.46mm) with MFCCs, DWCs, and SCs features. The CP stop sounds are better reconstructed using SPN (0.10mm) with SCs} \\
\hline
\small{\cite{Xie2016}} & \small{The proposed DNNs consist of 5 feedforward hidden layer with 512 units. The bottleneck features extracted from hierarchical inversion system used as auxiliary feature for tandem GVP-HMM inversion system} & \small{The hierarchical inversion systems (DNN - 2.588mm, RNN – 2.411mm) outperformed the baseline DNNs (2.633mm to 2.717mm) and RNNs (2.449mm to 2.493mm) inversion model} \\
\hline
\small{\cite{Wei2016}} & \small{Deep Auto-Encoder (DA) with 3 hidden layers (1000- 500-250 units) used to extract the features from ultrasound image and synchronized speech sound. The ultrasound to speech conversion performed using DNN} & \small{Mean Format Error (MFE), evaluates the predicted result with four formants of vowel, shows a promising result with 5\% betterment (F1-3.16\%, F2- 4.30\%, F3-1.13\%, and F4-2.31\%)} \\
\hline
\small{\cite{Sivaraman2016}} & \small{GMM used for acoustic modelling. VTLN used to normalize multiple speakers. The inversion system trained using feed-forward NN and Kalman smoothing technique used for predicted TV smoothing} & \small{The speech inversion system with VTLN provides an average correlation improvement of 7\% over sys1 (0.55) to sys2 (0.62)} \\
\hline
\end{tabular}
\caption{Methods and outcome of the literature observed}
\label{tab:out}
\end{table*}
\section{Conclusion}
The Acoustic-to-Articulatory Inversion (AAI) is an
interesting and challenging research area for several decades.
AAI is a non-linear, data-driven regression problem which is
very specific on its data requirement. The acoustic and
articulatory feature representation is required to model AAI.
The model is trained with acoustic and corresponding
articulatory feature. The Mel Frequency Cepstral Coefficient
(MFCC), and its derivatives are the commonly used acoustic
features. The articulatory features are mostly extracted from
biomedical images such as EMA, UTI and so on. Large dataset
is available for speech only. But for parallel dataset collection
the requirements of the equipment, time consumption, and
uncomfortableness of attaching such equipment with the human
body are the main parameters that restrict from having a large
dataset. This study conducted on different data-driven
approaches that proved its efficiency in AAI. Now a days, the
availability of dataset (Table 2) facilitates the efficiency of AAI
in many applications. The application of Deep Neural Network
drastically improved the performance of articulatory
reconstruction. Recently, the studies escalated its applications
to the AAI of pathological subjects. The articulatory visual
feedback constructed from acoustic features can be used as an
aid for providing speech and language therapy or training. The
automated such system will be able to track the improvement
of each session, and can easily point out the articulatory
positions. This system will be equally useful for hearing
impaired individuals as they can view the feedback on screen.
But still, there exist many gaps that needs to be explored. An
innovative user-friendly 3D feedback system will be more
useful for the clinical context.

\bibliography{acl_latex}

\begin{thebibliography}{54}
\providecommand{\natexlab}[1]{#1}

\bibitem[{Afshan and Ghosh(2015)}]{Afshan2015}
A.~Afshan and P.~K. Ghosh. 2015.
\newblock Improved subject-independent acoustic-to-articulatory inversion.
\newblock \emph{Speech Communication}, 66:1--16.

\bibitem[{Ananthakrishnan and Engwall(2011)}]{Ananthakrishnan2011}
G.~Ananthakrishnan and O.~Engwall. 2011.
\newblock Mapping between acoustic and articulatory gestures.
\newblock \emph{Speech Communication}, 53(4):567--589.

\bibitem[{Atal et~al.(1978)Atal, Chang, Mathews, and Tukey}]{Atal1978}
B.~S. Atal, J.~J. Chang, M.~V. Mathews, and J.~W. Tukey. 1978.
\newblock Inversion of articulatory‐to‐acoustic transformation in the vocal tract by a computer‐sorting technique.
\newblock \emph{The Journal of the Acoustical Society of America}, 63(5):1535--1555.

\bibitem[{Baken(1992)}]{Baken1992}
R.~J. Baken. 1992.
\newblock Electroglottography.
\newblock \emph{Journal of Voice}, 6(2):98--110.

\bibitem[{Biasutto–Lervat and Ouni(2018)}]{Biasutto-Lervat2018}
T.~Biasutto–Lervat and S.~Ouni. 2018.
\newblock Phoneme-to-articulatory mapping using bidirectional gated rnn.
\newblock In \emph{Interspeech 2018-19th Annual Conference of the International Speech Communication Association}.

\bibitem[{Bozorg and Johnson(2020)}]{Bozorg2020}
N.~Bozorg and M.~T. Johnson. 2020.
\newblock Acoustic-to-articulatory inversion with deep autoregressive articulatory-wavenet.
\newblock \emph{Networks (CNNs)}, 22:23.

\bibitem[{Byun et~al.(2014)Byun, Hitchcock, and Swartz}]{Byun2014}
T.~M. Byun, E.~R. Hitchcock, and M.~T. Swartz. 2014.
\newblock Retroflex versus bunched in treatment for rhotic misarticulation: Evidence from ultrasound biofeedback intervention.
\newblock \emph{Journal of Speech, Language, and Hearing Research}, 57(6):2116--2130.

\bibitem[{Cai et~al.(2018)Cai, Qin, Cai, Li, Liu, and Zhong}]{Cai2018}
Z.~Cai, X.~Qin, D.~Cai, M.~Li, X.~Liu, and H.~Zhong. 2018.
\newblock The dku-jnu-ema electromagnetic articulography database on mandarin and chinese dialects with tandem feature based acoustic-to-articulatory inversion.
\newblock In \emph{2018 11th International Symposium on Chinese Spoken Language Processing (ISCSLP)}, pages 235--239. IEEE.

\bibitem[{Cleland et~al.(2015)Cleland, Scobbie, and Wrench}]{Cleland2015}
J.~Cleland, J.~M. Scobbie, and A.~A. Wrench. 2015.
\newblock Using ultrasound visual biofeedback to treat persistent primary speech sound disorders.
\newblock \emph{Clinical linguistics \& phonetics}, 29(8-10):575--597.

\bibitem[{Fabre et~al.(2017)Fabre, Hueber, Girin, Alameda-Pineda, and Badin}]{Fabre2017}
D.~Fabre, T.~Hueber, L.~Girin, X.~Alameda-Pineda, and P.~Badin. 2017.
\newblock Automatic animation of an articulatory tongue model from ultrasound images of the vocal tract.
\newblock \emph{Speech Communication}, 93:63--75.

\bibitem[{Ghosh and Narayanan(2011)}]{Ghosh2011}
P.~K. Ghosh and S.~Narayanan. 2011.
\newblock Automatic speech recognition using articulatory features from subject-independent acoustic-to-articulatory inversion.
\newblock \emph{The Journal of the Acoustical Society of America}, 130(4):EL251--EL257.

\bibitem[{Giegerich(1992)}]{Giegerich1992}
H.~Giegerich. 1992.
\newblock \href {https://doi.org/10.1017/CBO9781139166126.002} {\emph{English Phonology: An Introduction}}.
\newblock Cambridge Textbooks in Linguistics. Cambridge University Press.

\bibitem[{Hardcastle(1972)}]{Hardcastle1972}
W.~J. Hardcastle. 1972.
\newblock The use of electropalatography in phonetic research.
\newblock \emph{Phonetica}, 25(4):197--215.

\bibitem[{Hogden et~al.(1996)Hogden, Lofqvist, Gracco, Zlokarnik, Rubin, and Saltzman}]{Hogden1996}
J.~Hogden, A.~Lofqvist, V.~Gracco, I.~Zlokarnik, P.~Rubin, and E.~Saltzman. 1996.
\newblock Accurate recovery of articulator positions from acoustics: New conclusions based on human data.
\newblock \emph{The Journal of the Acoustical Society of America}, 100(3):1819--1834.

\bibitem[{Hong et~al.(2020)Hong, Wu, Wang, and Huang}]{Hong2020}
Q.~B. Hong, C.~H. Wu, H.~M. Wang, and C.~L. Huang. 2020.
\newblock Combining deep embeddings of acoustic and articulatory features for speaker identification.
\newblock In \emph{ICASSP 2020-2020 IEEE International Conference on Acoustics, Speech and Signal Processing (ICASSP)}, pages 7589--7593. IEEE.

\bibitem[{Hueber et~al.(2013)Hueber, Bailly, Badin, and Elisei}]{Hueber2013}
T.~Hueber, G.~Bailly, P.~Badin, and F.~Elisei. 2013.
\newblock Speaker adaptation of an acoustic-to-articulatory inversion model using cascaded gaussian mixture regressions.
\newblock In \emph{Proceedings of the Annual Conference of the International Speech Communication Association, Interspeech}.

\bibitem[{Hueber et~al.(2012)Hueber, Ben-Youssef, Bailly, Badin, and Elisei}]{Hueber2012}
T.~Hueber, A.~Ben-Youssef, G.~Bailly, P.~Badin, and F.~Elisei. 2012.
\newblock Cross-speaker acoustic-to-articulatory inversion using phone-based trajectory hmm for pronunciation training.
\newblock In \emph{Thirteenth Annual Conference of the International Speech Communication Association}.

\bibitem[{Illa and Ghosh(2020)}]{Illa2020}
A.~Illa and P.~K. Ghosh. 2020.
\newblock The impact of speaking rate on acoustic-to-articulatory inversion.
\newblock \emph{Computer Speech \& Language}, 59:75--90.

\bibitem[{Jones(2017)}]{Jones2017}
D.~K. Jones. 2017.
\newblock \emph{Development of kinematic templates for automatic pronunciation assessment using acoustic-to-articulatory inversion}.
\newblock Doctoral dissertation, Marquette University.

\bibitem[{Kirchhoff et~al.(2002)Kirchhoff, Fink, and Sagerer}]{Kirchhoff2002}
K.~Kirchhoff, G.~A. Fink, and G.~Sagerer. 2002.
\newblock Combining acoustic and articulatory feature information for robust speech recognition.
\newblock \emph{Speech Communication}, 37(3-4):303--319.

\bibitem[{Krishna et~al.(2019)Krishna, Tran, Yu, and Tewfik}]{Krishna2019}
G.~Krishna, C.~Tran, J.~Yu, and A.~H. Tewfik. 2019.
\newblock Speech recognition with no speech or with noisy speech.
\newblock In \emph{ICASSP 2019-2019 IEEE International Conference on Acoustics, Speech and Signal Processing (ICASSP)}, pages 1090--1094. IEEE.

\bibitem[{Li et~al.(2019)Li, Chen, Siniscalchi, and Lee}]{Li2019}
W.~Li, N.~F. Chen, S.~M. Siniscalchi, and C.~H. Lee. 2019.
\newblock Improving mispronunciation detection of mandarin tones for non-native learners with soft-target tone labels and blstm-based deep tone models.
\newblock \emph{IEEE/ACM Transactions on Audio, Speech, and Language Processing}, 27(12):2012--2024.

\bibitem[{Ling et~al.(2009)Ling, Richmond, Yamagishi, and Wang}]{Ling2009}
Z.~H. Ling, K.~Richmond, J.~Yamagishi, and R.~H. Wang. 2009.
\newblock Integrating articulatory features into hmm-based parametric speech synthesis.
\newblock \emph{IEEE Transactions on Audio, Speech, and Language Processing}, 17(6):1171--1185.

\bibitem[{Liu et~al.(2015)Liu, Yu, Wu, Kang, Meng, and Cai}]{Liu2015}
P.~Liu, Q.~Yu, Z.~Wu, S.~Kang, H.~Meng, and L.~Cai. 2015.
\newblock A deep recurrent approach for acoustic-to-articulatory inversion.
\newblock In \emph{2015 IEEE International Conference on Acoustics, Speech and Signal Processing (ICASSP)}, pages 4450--4454. IEEE.

\bibitem[{Maharana et~al.(2021)Maharana, Illa, Mannem, Belur, Shetty, Kumar, and Ghosh}]{Maharana2021}
S.~K. Maharana, A.~Illa, R.~Mannem, Y.~Belur, P.~Shetty, V.~P. Kumar, and P.~K. Ghosh. 2021.
\newblock Acoustic-to-articulatory inversion for dysarthric speech by using cross-corpus acoustic-articulatory data.
\newblock In \emph{ICASSP 2021-2021 IEEE International Conference on Acoustics, Speech and Signal Processing (ICASSP)}, pages 6458--6462. IEEE.

\bibitem[{Mannem et~al.(2019)Mannem, Mallela, Illa, and Ghosh}]{Mannem2019}
R.~Mannem, J.~Mallela, A.~Illa, and P.~K. Ghosh. 2019.
\newblock Acoustic and articulatory feature based speech rate estimation using a convolutional dense neural network.
\newblock In \emph{Interspeech}, pages 929--933.

\bibitem[{Miller and Daniloff(1993)}]{Miller1993}
C.~J. Miller and R.~Daniloff. 1993.
\newblock Airflow measurements: theory and utility of findings.
\newblock \emph{Journal of Voice}, 7(1):38--46.

\bibitem[{Moro-Velazquez et~al.(2021)Moro-Velazquez, Gomez-Garcia, Arias-Londoño, Dehak, and Godino-Llorente}]{Moro-Velazquez2021}
L.~Moro-Velazquez, J.~A. Gomez-Garcia, J.~D. Arias-Londoño, N.~Dehak, and J.~I. Godino-Llorente. 2021.
\newblock Advances in parkinson's disease detection and assessment using voice and speech: A review of the articulatory and phonatory aspects.
\newblock \emph{Biomedical Signal Processing and Control}, 66:102418.

\bibitem[{Munhall et~al.(1995)Munhall, Vatikiotis‐Bateson, and Tohkura}]{Munhall1995}
K.~G. Munhall, E.~Vatikiotis‐Bateson, and Y.~Tohkura. 1995.
\newblock X‐ray film database for speech research.
\newblock \emph{The Journal of the Acoustical Society of America}, 98(2):1222--1224.

\bibitem[{Najnin and Banerjee(2015)}]{Najnin2015}
S.~Najnin and B.~Banerjee. 2015.
\newblock Improved speech inversion using general regression neural network.
\newblock \emph{The Journal of the Acoustical Society of America}, 138(3):EL229--EL235.

\bibitem[{Narayanan et~al.(2014)Narayanan, Toutios, Ramanarayanan, Lammert, Kim, Lee, and Proctor}]{Narayanan2014}
S.~Narayanan, A.~Toutios, V.~Ramanarayanan, A.~Lammert, J.~Kim, S.~Lee, and M.~Proctor. 2014.
\newblock Real-time magnetic resonance imaging and electromagnetic articulography database for speech production research (tc).
\newblock \emph{The Journal of the Acoustical Society of America}, 136(3):1307--1311.

\bibitem[{Ouni and Laprie(1999)}]{Ouni1999}
S.~Ouni and Y.~Laprie. 1999.
\newblock Design of hypercube codebooks for the acoustic-to-articulatory inversion respecting the non-linearities of the articulatory-to-acoustic mapping.
\newblock In \emph{Eurospeech}.

\bibitem[{Porras et~al.(2019)Porras, Sepúlveda-Sepúlveda, and Csapó}]{Porras2019}
D.~Porras, A.~Sepúlveda-Sepúlveda, and T.~G. Csapó. 2019.
\newblock Dnn-based acoustic-to-articulatory inversion using ultrasound tongue imaging.
\newblock In \emph{2019 International Joint Conference on Neural Networks (IJCNN)}, pages 1--8. IEEE.

\bibitem[{Richmond et~al.(2003)Richmond, King, and Taylor}]{Richmond2003}
K.~Richmond, S.~King, and P.~Taylor. 2003.
\newblock Modelling the uncertainty in recovering articulation from acoustics.
\newblock \emph{Computer Speech \& Language}, 17(2-3):153--172.

\bibitem[{Roxburgh et~al.(2015)Roxburgh, Scobbie, and Cleland}]{Roxburgh2015}
Z.~Roxburgh, J.~M. Scobbie, and J.~Cleland. 2015.
\newblock Articulation therapy for children with cleft palate using visual articulatory models and ultrasound biofeedback.
\newblock In \emph{Proceedings of the 18th International Congress of Phonetic Sciences (ICPhS), Glasgow, 10-14 August 2015}. International Phonetic Association.

\bibitem[{Rudzicz et~al.(2016)Rudzicz, Frydenlund, Robertson, and Thaine}]{Rudzicz2016}
F.~Rudzicz, A.~Frydenlund, S.~Robertson, and P.~Thaine. 2016.
\newblock Acoustic-articulatory relationships and inversion in sum-product and deep-belief networks.
\newblock \emph{Speech Communication}, 79:61--73.

\bibitem[{Ryalls and Behrens(2000)}]{Ryalls2000}
J.~H. Ryalls and S.~Behrens. 2000.
\newblock \emph{Introduction to speech science: From basic theories to clinical applications}.
\newblock Pearson College Division.

\bibitem[{Seneviratne and Espy-Wilson(2021)}]{Seneviratne2021}
N.~Seneviratne and C.~Espy-Wilson. 2021.
\newblock Speech based depression severity level classification using a multi-stage dilated cnn-lstm model.
\newblock \emph{arXiv preprint arXiv:2104.04195}.

\bibitem[{Seneviratne et~al.(2019)Seneviratne, Sivaraman, and Espy-Wilson}]{Seneviratne2019}
N.~Seneviratne, G.~Sivaraman, and C.~Y. Espy-Wilson. 2019.
\newblock Multi-corpus acoustic-to-articulatory speech inversion.
\newblock In \emph{INTERSPEECH}, pages 859--863.

\bibitem[{Sivaraman et~al.(2019)Sivaraman, Mitra, Nam, Tiede, and Espy-Wilson}]{Sivaraman2019}
G.~Sivaraman, V.~Mitra, H.~Nam, M.~Tiede, and C.~Espy-Wilson. 2019.
\newblock Unsupervised speaker adaptation for speaker independent acoustic to articulatory speech inversion.
\newblock \emph{The Journal of the Acoustical Society of America}, 146(1):316--329.

\bibitem[{Sivaraman et~al.(2016)Sivaraman, Mitra, Nam, Tiede, and Espy-Wilson}]{Sivaraman2016}
G.~Sivaraman, V.~Mitra, H.~Nam, M.~K. Tiede, and C.~Y. Espy-Wilson. 2016.
\newblock Vocal tract length normalization for speaker independent acoustic-to-articulatory speech inversion.
\newblock In \emph{INTERSPEECH}, pages 455--459.

\bibitem[{Steiner et~al.(2012)Steiner, Richmond, and Ouni}]{Steiner2012}
I.~Steiner, K.~Richmond, and S.~Ouni. 2012.
\newblock Using multimodal speech production data to evaluate articulatory animation for audiovisual speech synthesis.
\newblock In \emph{Proceedings of the 3rd Symposium on Facial Analysis and Animation}, pages 1--1.

\bibitem[{Sudhakar and Ghosh(2014)}]{Sudhakar2014}
P.~Sudhakar and P.~K. Ghosh. 2014.
\newblock Sparse smoothing of articulatory features from gaussian mixture model based acoustic-to-articulatory inversion: Benefit to speech recognition.
\newblock In \emph{Fifteenth Annual Conference of the International Speech Communication Association}.

\bibitem[{Sun et~al.(2021)Sun, Huang, Wang, and Zhang}]{Sun2021}
G.~Sun, Z.~Huang, L.~Wang, and P.~Zhang. 2021.
\newblock Temporal convolution network based joint optimization of acoustic-to-articulatory inversion.
\newblock \emph{Applied Sciences}, 11(19):9056.

\bibitem[{Suzuki et~al.(1998)Suzuki, Okadome, and Honda}]{Suzuki1998}
S.~Suzuki, T.~Okadome, and M.~Honda. 1998.
\newblock Determination of articulatory positions from speech acoustics by applying dynamic articulatory constraints.
\newblock In \emph{Fifth International Conference on Spoken Language Processing}.

\bibitem[{Tobing et~al.(2016)Tobing, Toda, Kameoka, and Nakamura}]{Tobing2016}
P.~L. Tobing, T.~Toda, H.~Kameoka, and S.~Nakamura. 2016.
\newblock Acoustic-to-articulatory inversion mapping based on latent trajectory gaussian mixture model.
\newblock In \emph{INTERSPEECH}, pages 953--957.

\bibitem[{Toda et~al.(2008)Toda, Black, and Tokuda}]{Toda2008}
T.~Toda, A.~W. Black, and K.~Tokuda. 2008.
\newblock Statistical mapping between articulatory movements and acoustic spectrum using a gaussian mixture model.
\newblock \emph{Speech Communication}, 50(3):215--227.

\bibitem[{Uchida et~al.(2017)Uchida, Saito, and Minematsu}]{Uchida2017}
H.~Uchida, D.~Saito, and N.~Minematsu. 2017.
\newblock Acoustic-to-articulatory mapping based on mixture of probabilistic canonical correlation analysis.
\newblock In \emph{INTERSPEECH}, pages 989--993.

\bibitem[{Uria et~al.(2012)Uria, Murray, Renals, and Richmond}]{Uria2012}
B.~Uria, I.~Murray, S.~Renals, and K.~Richmond. 2012.
\newblock Deep architectures for articulatory inversion.
\newblock In \emph{Thirteenth Annual Conference of the International Speech Communication Association}.

\bibitem[{Wei et~al.(2016)Wei, Fang, Zheng, Lu, He, and Dang}]{Wei2016}
J.~Wei, Q.~Fang, X.~Zheng, W.~Lu, Y.~He, and J.~Dang. 2016.
\newblock Mapping ultrasound-based articulatory images and vowel sounds with a deep neural network framework.
\newblock \emph{Multimedia Tools and Applications}, 75(9):5223--5245.

\bibitem[{Xie et~al.(2018)Xie, Liu, Lee, and Wang}]{Xie2018_isclsp}
X.~Xie, X.~Liu, T.~Lee, and L.~Wang. 2018.
\newblock Investigation of stacked deep neural networks and mixture density networks for acoustic-to-articulatory inversion.
\newblock In \emph{2018 11th International Symposium on Chinese Spoken Language Processing (ISCSLP)}, pages 36--40. IEEE.

\bibitem[{Xie et~al.(2016)Xie, Liu, and Wang}]{Xie2016}
X.~Xie, X.~Liu, and L.~Wang. 2016.
\newblock Deep neural network based acoustic-to-articulatory inversion using phone sequence information.
\newblock In \emph{Interspeech}, pages 1497--1501.

\bibitem[{Xie et~al.(2015)Xie, Liu, Wang, and Su}]{Xie2015}
X.~Xie, X.~Liu, L.~Wang, and R.~Su. 2015.
\newblock Generalized variable parameter hmms based acoustic-to-articulatory inversion.
\newblock In \emph{Sixteenth Annual Conference of the International Speech Communication Association}.

\bibitem[{Yu et~al.(2018)Yu, Yu, and Ling}]{Yu2018}
L.~Yu, J.~Yu, and Q.~Ling. 2018.
\newblock Synthesizing 3d acoustic-articulatory mapping trajectories: Predicting articulatory movements by long-term recurrent convolutional neural network.
\newblock In \emph{2018 IEEE Visual Communications and Image Processing (VCIP)}, pages 1--4. IEEE.

\end{thebibliography}

\end{document}